\documentclass[twocolumn,dviout]{aastex631}

\usepackage{natbib}
\usepackage{ulem}
\usepackage{xcolor}
\DeclareRobustCommand{\erase}{\bgroup\markoverwith{\textcolor{red}{\rule[.5ex]{2pt}{0.4pt}}}\ULon}

\shorttitle{Spatial extent of molecular gas, dust, and stars}
\shortauthors{Tadaki et al.}
\begin{document}

\title{Spatial extent of molecular gas, dust, and stars in massive galaxies at $z\sim2.2-2.5$ determined with ALMA and JWST}

\email{tadaki@hgu.jp}

\author[0000-0001-9728-8909]{Ken-ichi Tadaki}
\affiliation{Faculty of Engineering, Hokkai-Gakuen University, Toyohira-ku, Sapporo 062-8605, Japan}

\author{Tadayuki Kodama}
\affiliation{Astronomical Institute, Tohoku University, 6-3, Aramaki, Aoba, Sendai, Miyagi, 980-8578, Japan}

\author{Yusei Koyama}
\affiliation{Subaru Telescope, National Astronomical Observatory of Japan, 650 North A'ohoku Place, Hilo, HI 96720, USA}
\affiliation{Department of Astronomical Science, SOKENDAI (The Graduate University for Advanced Studies), Mitaka, Tokyo 181-8588, Japan}

\author{Tomoko L. Suzuki}
\affiliation{Kavli Institute for the Physics and Mathematics of the Universe (WPI),The University of Tokyo Institutes for Advanced Study, The University of Tokyo, Kashiwa, Chiba 277-8583, Japan}

\author{Ikki Mitsuhashi}
\affiliation{National Astronomical Observatory of Japan, 2-21-1 Osawa, Mitaka, Tokyo 181-8588, Japan}
\affiliation{Department of Astronomy, The University of Tokyo, 7-3-1 Hongo, Bunkyo, Tokyo 113-0033, Japan}

\author[0000-0002-2634-9169]{Ryota Ikeda}
\affiliation{Department of Astronomical Science, SOKENDAI (The Graduate University for Advanced Studies), Mitaka, Tokyo 181-8588, Japan}
\affiliation{National Astronomical Observatory of Japan, 2-21-1 Osawa, Mitaka, Tokyo 181-8588, Japan}



\begin{abstract}

We present the results of 0\farcs6-resolution observations of CO $J=3-2$ line emission in 10 massive star-forming galaxies at $z\sim2.2-2.5$ with the Atacama Large Millimeter/submillimeter Array (ALMA).
We compare the spatial extent of molecular gas with those of dust and stars, traced by the 870 $\mu$m and 4.4 $\mu$m continuum emissions, respectively.
The average effective radius of the CO emission is 1.75$\pm$0.34 kpc, which is about 60 percent larger than that of the 870 $\mu$m emission and is comparable with that of the 4.4 $\mu$m emission.
Utilizing the best-fit parametric models, we derive the radial gradients of the specific star-formation rate (sSFR), gas depletion timescale, and gas-mass fraction within the observed galaxies.
We find a more intense star-formation activity with a higher sSFR and a shorter depletion timescale in the inner region than in the outer region.
The central starburst may be the primary process for massive galaxies to build up a core.
Furthermore, the gas-mass fraction is high, independent of the galactocentric radius in the observed galaxies, suggesting that the galaxies have not begun to quench star formation.
Given the shorter gas depletion timescale in the center compared to the outer region, quenching is expected to occur in the center first and then propagate outward.
We may be witnessing the observed galaxies in the formation phase of a core prior to the forthcoming phase of star formation propagating outward.

\end{abstract}

\keywords{galaxies: starburst --- galaxies: high-redshift --- galaxies: ISM}


\section{Introduction} \label{sec:intro}

Massive galaxies populate two distinct regions on the stellar mass vs. star-formation rate (SFR) plane: the so-called main-sequence of star-forming galaxies and quiescent regime below the main-sequence \citep[e.g.][]{2015ApJ...801L..29R}.
At $z=1-3$, quiescent galaxies have a compact core in the center \citep[e.g.][]{2007ApJ...671..285T} and their radial profile of the stellar emission is well characterized by S$\acute{\mathrm{e}}$rsic models with a S$\acute{\mathrm{e}}$rsic index of $n=4$ \citep[e.g.][]{2005ApJ...626..680D}. In contrast, star-forming galaxies (SFGs) at similar redshifts have an exponential disk with $n=1$ \citep[e.g.][]{2011ApJ...742...96W}.
The bimodality on the stellar mass--SFR plane and the correlation between star-formation activity and morphology suggest that quenching of the SFR and transformation of galaxy morphology occur at almost the same epoch and on a short timescale.

The scenario of the star formation spreading outward from the galaxy center (so-called ``inside-out growth'') is widely accepted to explain the morphological evolution of massive galaxies \citep{2010ApJ...709.1018V}, i.e.,
galaxies form a compact core and later build the outer envelope through star formation and minor mergers \citep{2009ApJ...697.1290B}.
ALMA observations have revealed that massive SFGs commonly have a dusty star-forming core with an effective radius of $R_\mathrm{e}$=1--2 kpc \citep[e.g.][]{2016ApJ...827L..32B,2018A&A...616A.110E}.
However, the spatial distributions of HST/1.6 $\mu$m emission suggest that the massive SFGs already have an extended stellar disk with $R_\mathrm{e}$=2--6 kpc \citep[hereafter T20]{2020ApJ...901...74T}.
These results do not necessarily support that massive SFGs are in the inside-out growth phase.

A difficulty to verify the inside-out scenario is that the 1.6 $\mu$m emission, 0.5 $\mu$m in the rest frame at $z\sim2$, can be attenuated by dust, particularly in the central region of the galaxy, where compact 870 $\mu$m emission is detected.
This dust-extinction effect leads to a situation where the spatial distribution of 1.6 $\mu$m emission is more extended than that of the stellar mass, which is predicted by numerical simulations with radiative transfer calculations \citep{2022MNRAS.510.3321P} and has been observationally confirmed with spectral energy distribution (SED) fitting approaches \citep{2019ApJ...877..103S}.
High-resolution 4.4 $\mu$m images obtained by JWST/NIRCam \citep{2023PASP..135b8001R} provide a more accurate view of galaxy morphology with less effect of dust extinction \citep{2022ApJ...937L..33S, 2023arXiv230703264V}.

Another problem is that the information about the spatial distribution of molecular gas is lacking, which makes it difficult for us to understand how massive SFGs form stars and quench star formation.
As spatially-resolved observations of distant galaxies with CO emission lines tracing molecular gas are still limited in normal SFGs in the main-sequence due to faint emission \citep{2017ApJ...846..108C,2020ApJ...899...37K,2022ApJ...933...11I,2023ApJ...942...98L}, no systematic surveys of the CO emission lines of galaxies coordinated with JWST 4.4 $\mu$m observations have ever been conducted.
In this letter, we present the results of ALMA observations of CO emission lines from 10 massive SFGs with a stellar mass of $\log(M_\star/M_\odot) > 11$ at $z\sim2.2-2.5$ in the PRIMER-UDS field (JWST Cycle 1 GO program; ID: 1837; PI: J. Dunlop) where JWST 4.4 $\mu$m images are obtained.
For understanding how massive SFGs form a core, we compare the spatial extent of three components: gas, dust, and stars.

\section{Observations}
\label{sec;obs}

\begin{figure}[!t]
\centering
\includegraphics[scale=1.0]{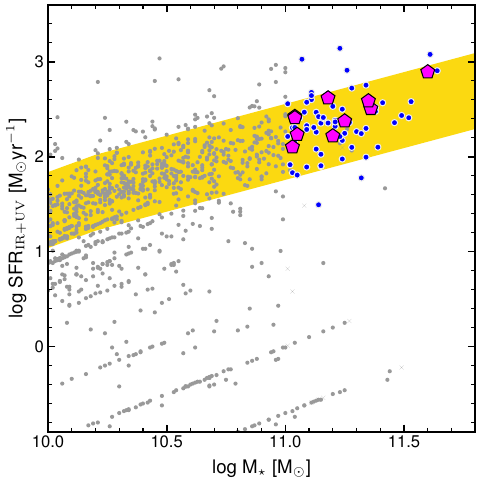}
\caption{
Stellar mass vs. SFR for our ALMA sample of 10 massive SFGs at $z\sim2.2-2.5$ (magenta pentagons), the T20 sample with ALMA 870 $\mu$m data (blue circles), and all galaxies at $1.9<z<2.6$ (gray dots) taken from the 3D-HST catalog (\citealt{2014ApJS..214...24S}; \citealt{2016ApJS..225...27M}). 
The shaded region indicates the range of $\pm$0.4 dex from the star-formation main sequence at $z\sim2$ \citep{2014ApJ...795..104W}.
}
\label{fig;MS}
\end{figure}

T20 reported ALMA 0\farcs2-resolution observations of 870 $\mu$m continuum emission in 62 massive SFGs at $1.9<z<2.6$ to characterize the spatial extent of dust emission.
In T20, the stellar masses were estimated with SED fitting of multi-wavelength photometry data from the ultraviolet to near-infrared of sources in the 3D-HST catalog \citep{2014ApJS..214...24S}. 
The total infrared luminosities were estimated from the single-band photometry at Herschel/PACS 160 $\mu$m, 100 $\mu$m, or Spitzer/MIPS 24 $\mu$m \citep{2011ApJ...738..106W}, and were converted to dust-obscured SFRs \citep{1998ARA&A..36..189K}.
Unobscured SFRs were estimated from the rest-frame 2800 \AA~luminosity, but they do not contribute little to total SFRs (T20).
Figure \ref{fig;MS} shows a scatter plot of the stellar mass vs. SFR for galaxies at $1.9<z<2.6$.
The T20 sample lies on the main sequence of star formation at $z\sim2$, suggesting that they are representative of massive SFGs at $z\sim2$ \citep{2014ApJ...795..104W}. 

\begin{figure*}[!]
\centering
\includegraphics[scale=1.0]{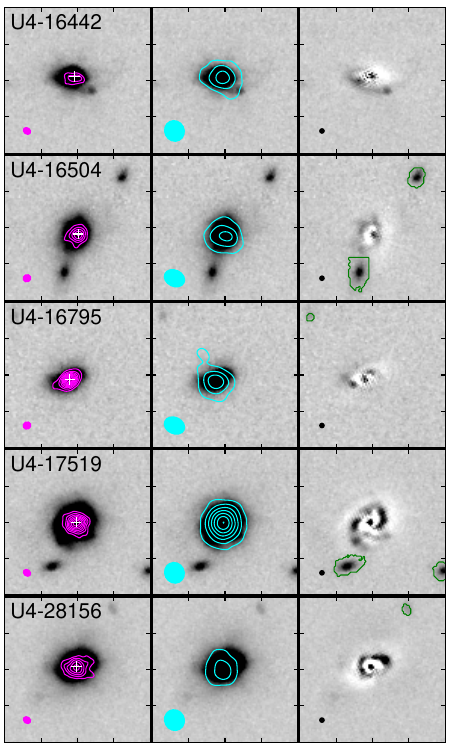}
\ \ \ \
\includegraphics[scale=1.0]{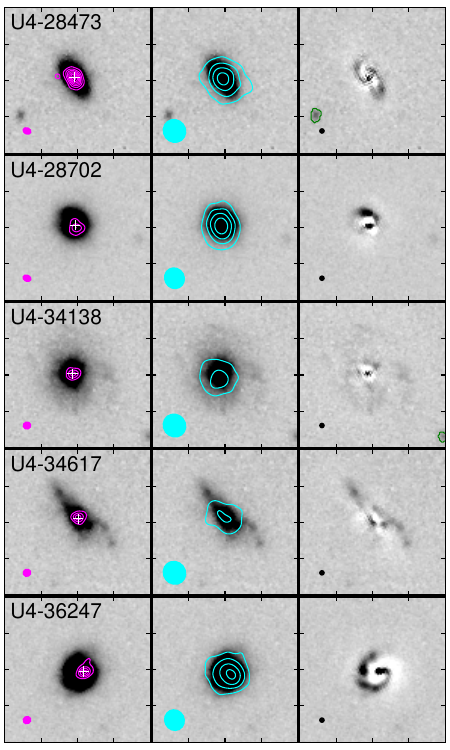}
\caption{
The JWST 4.4 $\mu$m images (4\arcsec$\times$4\arcsec) for our sample of 10 massive SFGs (left and center columns in each panel).
Magenta contours display the 870 $\mu$m flux densities in the ALMA 0\farcs2-resolution images, plotted for every 4$\sigma$.
Cyan contours do the velocity-integrated flux of CO $J=3-2$ emission in the same image, plotted for every 4$\sigma$.
The right columns in each panel show the residual 4.4 $\mu$m images after the best-fit model is subtracted.
Magenta, cyan, and black filled ellipses at the bottom-left corner correspond to the FWHM of PSF in the 870 $\mu$m, CO, and 4.4 $\mu$m images, respectively.
Green lines show the regions masked in measuring the effective radii of the 4.4 $\mu$m emission.
White crosses show the central position of the best-fit model in the JWST 4.4 $\mu$m image.
}
\label{fig;image}
\end{figure*}

For two (U4-16504 and U4-16795) of the 62 massive SFGs, the CO $J=3-2$ line emission has been observed with ALMA at a 0\farcs6 resolution \citep[hereafter T17]{2017ApJ...841L..25T}. 
From the T20 sample, we select additional six massive SFGs in redshift ranges of $z=2.17-2.21$ and $z=2.51-2.55$.
Their CO $J=3-2$ emission lines can be covered with a single frequency setup of ALMA Band-3 receivers since the redshift is accurately determined through H$\alpha$ spectroscopy ($\Delta z < 0.01$; \citealt{2019ApJ...886..124W}) or H$\alpha$ narrow-band imaging ($\Delta z < 0.02$; \citealt{2013ApJ...778..114T}).
ALMA observations targeting the six massive SFGs were conducted in 2021 December and 2022 July. 
The on-source time was 85 minutes per pointing. 
The data were calibrated in the standard manner using {\tt CASA} \citep{2022PASP..134k4501C}.
Then, we cleaned the emission down to the 1.5$\sigma$ level to create 50 km s$^{-1}$ channel maps with a robust parameter of +2.0, leading to a spatial resolution of 0\farcs6 and a noise level of 0.13--0.14 mJy beam$^{-1}$.
The data quality is similar to that in T17.
When the line emission is integrated over a velocity range of 200--750 km s$^{-1}$,
the CO emission is detected at more than 8$\sigma$ for all the targets (Figure \ref{fig;image}).
In addition, two massive SFGs (U4-17519 and U4-28156) of the T20 sample are serendipitously detected in CO in the same ALMA observations.
Adding two from T17, we have obtained a sample of 10 massive SFGs in total (Table \ref{tab;table}).
We extract the CO spectra in an aperture of 1\arcsec diameter and fit them to Gaussian profiles to determine the spectroscopic redshift (Table \ref{tab;table}).
Our sample is the largest sample of massive SFGs at $z\sim2.2-2.5$ with spatially-resolved data of both 870 $\mu$m continuum and CO $J=3-2$ line emission.

\begin{table*}
\caption{ Source list of 10 massive SFGs \label{tab;table}}
\begin{center}
\begin{tabular}{lcccccccc}
\hline
ID  & $z_\mathrm{CO}$  & $R_\mathrm{e,870}$\tablenotemark{a} & $q_\mathrm{870}$\tablenotemark{b} & $Sdv_\mathrm{CO}$\tablenotemark{c} & $R_\mathrm{e,CO}$\tablenotemark{a} & $R_\mathrm{e,4.4}$ \tablenotemark{a,e} & $q_\mathrm{4.4}$\tablenotemark{b,e} & $n_\mathrm{4.4}$\tablenotemark{d,e} \\
 & & (kpc) & & (Jy km s$^{-1}$) & (kpc) & (kpc) & &\\
\hline
U4-16442 & 2.217$\pm$0.003 & 0.97$\pm$0.14 & 0.29$\pm$0.12 & 0.70$\pm$0.08 & 1.33$\pm$0.30 & 0.85 & 0.50 & 3.2 \\
U4-16504 & 2.527$\pm$0.003 & 1.25$\pm$0.11 & 0.85$\pm$0.14 & 0.82$\pm$0.11 & 2.30$\pm$0.42 & 2.12 & 0.78 & 1.7 \\
U4-16795 & 2.523$\pm$0.004 & 0.91$\pm$0.05 & 0.53$\pm$0.06 & 0.53$\pm$0.08 & 1.34$\pm$0.35 & 1.40 & 0.48 & 1.2 \\
U4-17519 & 2.222$\pm$0.001 & 1.56$\pm$0.06 & 0.92$\pm$0.06 & 1.61$\pm$0.10 & 2.00$\pm$0.20 & 2.06 & 0.86 & 1.3 \\
U4-28156 & 2.209$\pm$0.004 & 1.73$\pm$0.08 & 0.72$\pm$0.06 & 0.89$\pm$0.15 & 1.48$\pm$0.49 & 1.74 & 0.68 & 1.4 \\
U4-28473 & 2.523$\pm$0.004 & 0.86$\pm$0.06 & 0.44$\pm$0.06 & 0.79$\pm$0.08 & 1.56$\pm$0.30 & 1.37 & 0.55 & 1.7 \\
U4-28702 & 2.176$\pm$0.001 & 1.02$\pm$0.16 & 0.67$\pm$0.23 & 0.51$\pm$0.05 & 2.01$\pm$0.32 & 1.65 & 0.95 & 1.3 \\
U4-34138 & 2.518$\pm$0.003 & 0.48$\pm$0.10 & 0.47$\pm$0.25 & 0.34$\pm$0.06 & 1.52$\pm$0.51 & 3.30 & 0.89 & 2.9 \\
U4-34617 & 2.533$\pm$0.004 & 0.30$\pm$0.07 & 0.27$\pm$0.33 & 0.30$\pm$0.06 & 1.42$\pm$0.55 & 2.69 & 0.44 & 1.6 \\
U4-36247 & 2.179$\pm$0.001 & 0.55$\pm$0.08 & 0.55$\pm$0.20 & 0.57$\pm$0.06 & 1.97$\pm$0.34 & 2.24 & 0.89 & 1.3 \\
\hline
\end{tabular}
\end{center}
\tablenotetext{a}{The circularized effective radius from the S$\acute{\mathrm{e}}$rsic model.}
\tablenotetext{b}{The major-to-minor axis ratio from the S$\acute{\mathrm{e}}$rsic model.}
\tablenotetext{c}{The line flux from the S$\acute{\mathrm{e}}$rsic model.}
\tablenotetext{d}{The S$\acute{\mathrm{e}}$rsic index from the S$\acute{\mathrm{e}}$rsic model.}
\tablenotetext{e}{The fitting errors are less than 1\%.}
\end{table*}

\section{Size measurements}
\label{sec;size_measurement}

\subsection{ALMA data of 870 $\mu$m and CO emissions}

CO and 870 $\mu$m continuum emissions were observed with an interferometer; then, we model the visibility data, not the images (Figure \ref{fig;image}). 
This approach has advantages of not being affected by uncertainties in Fourier transforms and deconvolution of a dirty beam.
For the 870 $\mu$m continuum emission, T20 have measured the flux density, effective radius, major-to-minor axis ratio, and position angle of each galaxy, assuming an elliptical exponential disk model, and characterized its spatial extent (Table \ref{tab;table}).
For U4-34138 and U4-34617, the best-fit models account for only 50\%–70\% of the total flux densities directly measured from the short-baseline data, suggesting that a single component model is not sufficient for characterizing the spatial distribution of the 870 $\mu$m continuum emission.
We do not use these two galaxies in section \ref{sec;comparison} and section \ref{sec;profile} as their effective radii can be underestimated.

In the same way as T17, we measure the effective radius of CO emission for 10 massive SFGs with model fitting.
In the fitting, we fix the axis ratio and the position angle of the CO emission to those of 870 $\mu$m emission because the spatial resolution of the CO observations is not as high as that of the 870 $\mu$m observations.
We fit elliptical exponential models to the velocity-integrated CO data, using the {\tt UVMULTIFIT} \citep{2014A&A...563A.136M}. 
Consequently, the ``circularized'' effective radius of CO emission is found to range from 1.3 kpc to 2.3 kpc (Table \ref{tab;table}).
Here, the circularized effective radius, $R_\mathrm{e}$, is obtained by multiplying the effective radius along the major axis by the square root of the axis ratio and is used in the following sections.

\subsection{JWST data of 4.4 $\mu$m emission}
\label{sec;JWSTsize}

Our sample of 10 massive SFGs are all located in the PRIMER-UDS field.
We use the v7 data release of mosaic images in the F444W filter (4.4 $\mu$m) with a pixel scale of 0\farcs04 from the Dawn JWST Archive, in which data were processed with the {\tt Grizli} software \citep{2023zndo...8210732B, 2023ApJ...947...20V}.
We measure the effective radii of the 4.4 $\mu$m continuum emission for massive SFGs in the following procedure.

First, we select 18 unsaturated stars with AB magnitudes of 20--21 in the Spitzer/IRAC 4.5 $\mu$m band \citep{2013ApJ...769...80A} from the 3D-HST catalog \citep{2014ApJS..214...24S} and stack their normalized cut-out images in a sub-pixel scale of 0\farcs02.
We use the stacked image for deconvolution of the point-spread function (PSF) in the following size measurements.
We generate segmentation maps of stars and massive SFGs by using {\tt SExtractor} \citep{1996A&AS..117..393B} to mask neighboring sources.  
The publicly released weight maps incorporate pixel-to-pixel variations and the noise from the sky background, but do not include the Poisson noise for individual sources.
We therefore make full variance images based on the original count-rate data and convert them to sigma images by following the procedure described in the Dawn JWST Archive\footnote{{https://dawn-cph.github.io/dja/blog/2023/07/18/image-data-products/}}.
Using the {\tt GALFIT} code \citep{2010AJ....139.2097P}, we fit JWST 4.4 $\mu$m cut-out images (4\arcsec$\times$4\arcsec) of 18 stars as a point source.
Four of 18 stars show the reduced chi-square value of $\chi_\nu^2>2$ or a companion in the unmasked region. 
Removing these four stars, we remake the stacked PSF image from the remaining 14 stars.
A full width at half maximum (FWHM) of a star in the updated stacked PSF image is 0\farcs16.

Next, we fit the cut-out images of massive SFGs with S$\acute{\mathrm{e}}$rsic models, fixing the sky value at zero and allowing seven parameters to vary: the centroid position, flux density, effective radius, S$\acute{\mathrm{e}}$rsic index, axis ratio, and position angle.
The derived centroid position of the 4.4 $\mu$m emission coincides well with the peak of 870 $\mu$m emission (Figure \ref{fig;image}).
The circularized effective radius is found to range from 0.8 kpc to 3.3 kpc (Table \ref{tab;table}).
The median value of the S$\acute{\mathrm{e}}$rsic index is $n=1.76$, which is intermediate between SFGs and quiescent galaxies \citep{2011ApJ...742...96W}.

The residual images after subtracting the best-fit model from the original images show sub-structures, such as spiral arms and off-center clumps (Figure \ref{fig;image}), which cannot be characterized by simple elliptical models.
Despite the presence of residual emissions, the fitting errors are very small, $\sim$1\% or less in each parameter.
This may be attributed to the fact that our sample sources are very bright at 4.4 $\mu$m with AB magnitudes of 20--21.5 and that the signal-to-noise ratio is high. The actual uncertainty is dominated by deviations from a simple parametric model, rather than the statistical errors of the fitting.

To verify the validity of S$\acute{\mathrm{e}}$rsic models, we measure the total flux densities in the JWST 4.4 $\mu$m images by using {\tt SExtractor/MAG\_ISOCOR} and compare them with the flux densities of the best-fit models.
We find that the models account for 98--107 \% of the total flux densities, suggesting that the spatial distribution of the primary component is mostly characterized by the S$\acute{\mathrm{e}}$rsic models.
An exception is U4-34138; the best-fit model accounts for 114\% of its total flux density, and
both ALMA 870 $\mu$m and JWST 4.4 $\mu$m indicate that a single component is not sufficient for explaining their spatial distributions.

\begin{figure*}[!t]
\centering
\includegraphics[scale=1.0]{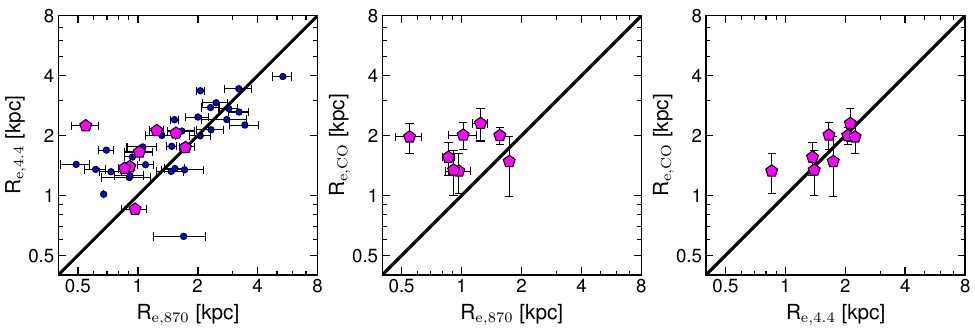}
\caption{
Size comparisons between the CO, 870 $\mu$m, and 4.4 $\mu$m emissions for eight massive SFGs (magenta pentagons).
The solid lines show the 1:1 correspondence between the circularized effective radii compared.
Blue circles indicate the T20 sample with ALMA 870 $\mu$m and JWST 4.4 $\mu$m data, but without CO data.
}
\label{fig;size}
\end{figure*}

\subsection{Size comparisons}
\label{sec;comparison}

For eight massive SFGs excluding U4-34138 and U4-34617, we have measured the effective radii of the CO, 870 $\mu$m, and 4.4 $\mu$m emissions.
The average and standard deviation of the effective radii of the CO, 870 $\mu$m, and 4.4 $\mu$m emissions are 1.75$\pm$0.34 kpc, 1.11$\pm$0.36 kpc, and 1.68$\pm$0.43 kpc, respectively.
Figure \ref{fig;size} shows size comparisons between the three tracers.
The effective radius of the 870 $\mu$m emission is 34\% smaller than that of the 4.4 $\mu$m emission.
This trend is also reported in submillimeter bright galaxies \citep{2022ApJ...939L...7C}.
The size difference between 870 $\mu$m and 4.4 $\mu$m is much smaller than the difference between 870 $\mu$m and 1.6 $\mu$m (T20), which is consistent with the fact that 1.6 $\mu$m should be much more severely affected by dust extinction.
We also measure the effective radius of 4.4 $\mu$m emission for additional 31 massive SFGs in the PRIMER-UDS field, selected from T20 sample with 870 $\mu$m size measurements, in the same way as section \ref{sec;JWSTsize}. 
We confirm that the 870 $\mu$m emission is more compact than the 4.4 $\mu$m emission in the range of $R_\mathrm{e,870}<2$ kpc (Figure \ref{fig;size}).
On the other hand, there appears to be little difference in size for large galaxies with $R_\mathrm{e,870}>2$ kpc.
We note that our sample in the CO observations is somewhat biased toward galaxies with compact dust emission.

We compare the effective radii of the CO emission with those of the 870 $\mu$m and 4.4 $\mu$m emissions.
The CO emission is more extended than the 870 $\mu$m emission, which is consistent with previous studies of galaxies at $z=1-3$ \citep{2017ApJ...846..108C,2018ApJ...863...56C,2022ApJ...933...11I}. 
Whereas the low-$J$ CO line emission traces the mass of interstellar medium, the 870 $\mu$m emission is sensitive to dust heating by massive stars associated with ongoing star-forming activity.
We discuss the size difference between molecular gas and dust in section \ref{sec;profile}. 
In contrast, the spatial extent of the CO emission is comparable with that of the 4.4 $\mu$m emission.
Although the current resolution and sensitivity of the CO data do not allow us to capture the sub-structures as seen in the JWST 4.4 $\mu$m images, the spatial distributions of the CO and 4.4 $\mu$m emissions may be the same, at least in a kilo-parsec scale.

\section{Radial profiles in massive galaxies}
\label{sec;profile}

CO, 870 $\mu$m, and 4.4 $\mu$m emissions trace molecular gas mass, dust-obscured SFR, and stellar mass, respectively.
We have characterized the spatial extent of each component and derived the best-fit models in section \ref{sec;size_measurement}.
Here we interpret these parametric models to investigate the radial dependence of physical properties in massive SFGs with a compact dusty core.
For the CO data, we estimate the radial profile of molecular gas masses from the CO $J=3-2$ luminosities, following the recipe of \citet{2020ARA&A..58..157T} for SFGs in the main sequence where
two conversion factors are used: $R_{31}$=1.8 for CO $J=3-2$ to CO $J=1-0$ and  $\alpha_\mathrm{CO}=4.36$ for CO to H$_2$.
For the 870 $\mu$m and 4.4 $\mu$m data, we do not directly convert their observed luminosities to the SFR and stellar mass, respectively.
Galaxy-integrated values of dust-obscured SFRs and stellar masses of the massive SFGs have been estimated from multi-wavelength data (T20).
By scaling the best-fit models of the 870 $\mu$m and 4.4 $\mu$m emissions to the galaxy-integrated values,
we obtain the radial profiles of the SFR and stellar mass, respectively.
In this work, we do not take into account for dust-unobscured star formation in the radial profile of SFR.
The spatial spatial distribution of H$\alpha$ emission, another tracer of star formation, is more extended than that of dust continuum emission in massive SFGs \citep{2020ApJ...892....1W, 2020ApJ...901...74T}, suggesting that the unobscured component dominates star formation in the outer region.
However, since the H$\alpha$-based SFR corresponds to only 10\% of the total SFR in massive SFGs \citep{2015ApJ...813...23V},
including the H$\alpha$ component could not significantly affect the estimate of the half SFR radii.

\begin{figure*}[!t]
\centering
\includegraphics[scale=1.0]{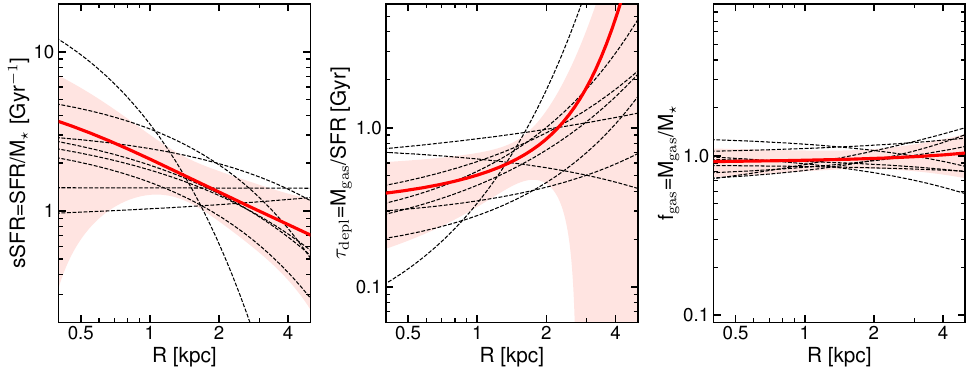}
\caption{
Best-fit model radial profiles of sSFR, gas depletion timescale, and gas mass fraction.
Solid red lines and shaded region show the average and standard deviation in eight massive SFGs, respectively.
Black dashed lines show the individual models.
}
\label{fig;profile}
\end{figure*}

Then, we calculate the radial profiles of the specific SFR (sSFR), defined as the SFR per unit stellar mass for eight massive SFGs (Figure \ref{fig;profile}). 
In the ``inside-out growth'' scenario (see Section~1), a core forms first in the massive galaxy and then star formation propagates outward.
In this phase, the sSFR is larger in the outer region than in the inner region because ongoing star-formation activity contributes to forming an extended disk.
Our results, however, show that the sSFR increases towards the center, which is the opposite trend reported in previous studies of massive galaxies \citep{2016ApJ...828...27N,2018ApJ...859...56T,2019ApJ...883...81S}.
If the current star formation is maintained, the radial profile of stellar mass will be more centrally concentrated.
This is supported by the fact that the S$\acute{\mathrm{e}}$rsic index of massive SFGs is higher than $n=1$ observed in less massive SFGs.
The massive SFGs may be building up a central core and transforming their morphology from disk-dominated to bulge-dominated, eventually leading to a $n=4$ profile as seen in massive quiescent galaxies at similar redshifts \citep{2011ApJ...742...96W,2014ApJ...788...11L}.

Moreover, we find that the central region has a shorter gas-depletion timescale $\tau_\mathrm{depl}$, defined as the gas mass divided by the SFR, than the outer region. 
The gas-depletion timescales in starburst galaxies are 0.5 dex shorter than those in normal SFGs, even when the same CO-to-H$_2$ conversion factor is used \citep{2021ApJ...908...61K}.
In normal SFGs, a radial dependence of $\tau_\mathrm{depl}$ is not seen in the regime where molecular gas dominates interstellar medium \citep{2008AJ....136.2782L}.
These results suggest the existence of a bimodality in star-formation activity, the normal mode and starburst mode, also as pointed out by studies of high-redshift galaxies \citep{2010ApJ...714L.118D, 2010MNRAS.407.2091G}.
The physical condition of gas in the central region of massive SFGs is likely similar to that in starburst galaxies, in which galaxy-galaxy mergers could trigger intense star formation.
Our sample of massive SFGs show no indication of major mergers, but some of them have potential, small companions within $\sim10$ kpc in their JWST 4.4 $\mu$m images (Figure \ref{fig;image}).
The tidal interaction by small satellites can develop non-axisymmetric structures in the galaxy, such as spiral arms, which drive gas flows to the center \citep{1994ApJ...425L..13M}.
The spiral-arm-driven starburst scenario is supported by the fact that non-axisymmetric substructures are identified in the JWST 4.4 $\mu$m images after the primary component is subtracted (Figure \ref{fig;image}).

The radial profile of gas-mass fraction $f_\mathrm{gas}$, defined by the gas mass divided by the stellar mass, tells where galaxies begin to quench.
The gas mass fraction is constant over the galaxy at a large gas fraction of about 1.0, indicating that galaxies are gas-rich over the entire region.
Even if a small CO-to-H$_2$ conversion factor of $\alpha_\mathrm{CO}=1$ in the starburst mode is applied, the gas mass fraction is still larger than that in nearby SFGs \citep{2017ApJS..233...22S}.
The presence of such large gas reservoirs implies that quenching has not begun in the massive SFGs.
If cold gas is not accreted onto galaxies from their surroundings, the existing gas begins to be consumed from the center by star formation within several hundred million years.
Our sample of massive SFGs are likely to be in the phase immediately before ``inside-out'' quenching \citep{2015Sci...348..314T}. 

Finally, we note a few caveats in our analysis.
Our approach implicitly assumes that luminosity-to-mass ratios are constant within a galaxy.
We use 4.4 $\mu$m emissions, which corresponds to the rest-frame 1.5 $\mu$m, as the tracer of the spatial distribution of the stellar mass.
The rest-frame 1.5 $\mu$m emission is less affected by dust extinction.
Nevertheless, \citet{2023MNRAS.524.4128Z} show, using radiative transfer models, that even 4.4 $\mu$m sizes are about 75 percent overestimated than the half mass radii in massive SFGs. 
If this effect is significant in our sample, the true spatial extent of stars is more compact than that of the SFR, indicating a suppressed sSFR in the center.
Also, a radial gradient of dust temperature affects the conversion of the 870 $\mu$m flux density to the total infrared luminosity or SFR.
When a dust temperature is higher at the center, SFR profiles could be more centrally concentrated than the 870 $\mu$m ones.
A deviation from parametric models is another problem to accurately characterize the spatial distribution of emissions.
Unlike the JWST 4.4 $\mu$m images, sub-structures are not visible in the ALMA 870 $\mu$m images.
This may simply be due to the low signal-to-noise ratio ($\sim$10--30) in the ALMA observations.
The peak flux density in the residual images of JWST 4.4 $\mu$m emission is less than 15\% of the peak in the original image (Figure \ref{fig;image}).
To capture the sub-structures of dust emission, it is necessary to have deeper 870 $\mu$m images than used in this work, as demonstrated in previous studies of bright galaxies \citep[e.g.,][]{2016ApJ...829L..10I, 2019ApJ...876..130H}.
Although these uncertainties are desired to be addressed in the future, we are beginning to see with excellent data from the JWST and ALMA how massive galaxies form a core.

\vspace{0.5cm}

We thank the referee for constructive comments that improved the paper. 
This paper makes use of the following ALMA data: ADS/JAO.ALMA\#2021.1.01291.S. ALMA is a partnership of ESO (representing its member states), NSF (USA) and NINS (Japan), together with NRC (Canada), MOST and ASIAA (Taiwan), and KASI (Republic of Korea), in cooperation with the Republic of Chile. The Joint ALMA Observatory is operated by ESO, AUI/NRAO and NAOJ.
This work is based in part on observations made with the NASA/ESA/CSA James Webb Space Telescope. 
These observations are associated with program \#1837.
We acknowledge the PRIMER team for developing their observing program with a zero-exclusive-access period.
The data products presented herein were retrieved from the Dawn JWST Archive (DJA). DJA is an initiative of the Cosmic Dawn Center, which is funded by the Danish National Research Foundation under grant No. 140.
This work was supported by JSPS KAKENHI Grant Numbers 20K14526. %23K03466.
Data analysis was in part carried out on the Multi-wavelength Data Analysis System operated by the Astronomy Data Center (ADC), National Astronomical Observatory of Japan.


\software{CASA \citep{2022PASP..134k4501C},  
          UVMULTIFIT \citep{2014A&A...563A.136M},
          GALFIT \citep{2010AJ....139.2097P}, 
          Source Extractor \citep{1996A&AS..117..393B}
          }

\section*{Appendix}
\section*{Size measurements through visibility fitting}

\begin{figure*}[h]
\centering
\includegraphics[scale=1]{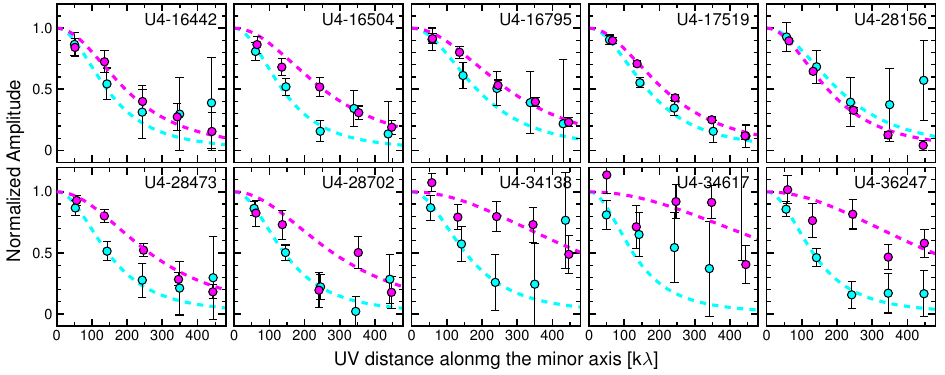}
\caption{
Normalized visibility amplitudes versus $uv$ distances for our sample of 10 massive SFGs. 
Magenta and cyan symbols show 870 $\mu$m and CO data, respectively.
Dashed lines indicate the best-fitting model for exponential profile.
}
\label{fig;uvamp}
\end{figure*}

We extract the spatial frequency ($u$,$v$) and the real/imaginary part of individual visibility by using {\tt CASA toolkit/table.getcol} to derive the $uv$ distance and the amplitudes.
We show in Figure \ref{fig;uvamp} the amplitudes of the visibilities as a function of the $uv$ distance along the minor axis for 870 $\mu$m and CO data.
For nine of 10 massive SFGs, the amplitudes of the CO data more rapidly decline as $uv$ distance compared to the 870 $\mu$m data, indicating that the CO emission is more compact.
We note that {\tt UVMULTIFIT} fits individual visibility data, not the averaged one shown in Figure \ref{fig;uvamp}.

\end{document}